\documentclass{article}
\usepackage{ijcai23}

\usepackage{times}
\usepackage{soul}
\usepackage{url}
\usepackage[hidelinks]{hyperref}
\usepackage[utf8]{inputenc}
\usepackage[small]{caption}
\usepackage{graphicx}
\usepackage{amsmath}
\usepackage{amsthm}
\usepackage{booktabs}
\usepackage{algorithm}
\usepackage{algorithmic}
\usepackage[switch]{lineno}
\usepackage{enumitem}
\usepackage{amssymb}

\title{MusicMagus: Zero-Shot Text-to-Music Editing via Diffusion Models}

\author{
Yixiao Zhang$^1$
\and
Yukara Ikemiya$^2$\and
Gus Xia$^{3}$\and
Naoki Murata$^2$\and \\
Marco A. Martínez-Ramírez$^2$\and 
Wei-Hsiang Liao$^2$\and
Yuki Mitsufuji$^2$\and
Simon Dixon$^1$
\affiliations
$^1$C4DM, Queen Mary University of London\\
$^2$Sony AI \\
$^3$Music X Lab, MBZUAI
\emails
\{yixiao.zhang, s.e.dixon\}@qmul.ac.uk,
\{yukara.ikemiya, naoki.murata, marco.martinez, weihsiang.liao, Yuhki.Mitsufuji\}@sony.com,
gus.xia@mbzuai.ac.ae}

\begin{document}

\maketitle

\begin{abstract}
    Recent advances in text-to-music generation models have opened new avenues in musical creativity. However, music generation usually involves iterative refinements, and how to edit the generated music remains a significant challenge. This paper introduces a novel approach to the editing of music generated by such models, enabling the modification of specific attributes, such as genre, mood and instrument, while maintaining other aspects unchanged. Our method transforms text editing to \textit{latent space manipulation} while adding an extra constraint to enforce consistency. It seamlessly integrates with existing pretrained text-to-music diffusion models without requiring additional training. Experimental results demonstrate superior performance over both zero-shot and certain supervised baselines in style and timbre transfer evaluations. We also showcase the practical applicability of our approach in real-world music editing scenarios. 
\end{abstract}

\section{Introduction}

Recent advances in text-to-music generation have unlocked new possibilities in musical creativity~\cite{butter,accomontage,musecoco,polyffusion,musiclm,mousai,musicgen,musicldm}. However, a significant challenge persists in how to \textit{edit} the generated results as music production usually involves iterative refinements. Building on this momentum, we regard `text-to-music editing' as the process of using text queries to edit music, and we see two major types of operations: \textit{inter-stem editing}, such as adding or removing instruments (e.g., ``add a saxophone" or ``remove the drums"), and \textit{intra-stem editing}, which involves modifications within the same stem, such as adding effects or changing instruments (e.g., ``add reverb to this stem" or ``transfer the timbre of the specified notes"). In this context, a ``stem'' refers to an individual track or component within a music piece, such as a specific instrument or vocal part. The primary focus of this paper is on the latter, \textit{intra-stem editing}. 

One of the fundamental challenges of text-to-music editing is the difficulty of accommodating flexible text operations in both dataset construction and model training. This is not only a matter of data pair scarcity, but also the complexity inherent in the vast array of possible text-based edits that can be applied to music. Existing research~\cite{audit,instructme,m2ugen} has primarily focused on manually constructing datasets. However, these models are constrained to a few predefined operations, which undermines their effectiveness in text-to-music editing that requires flexibility and variety. This highlights the need for a new approach that moves away from traditional supervised learning reliant on specific data pairs and towards a more adaptable, unsupervised, or zero-shot approach.

\begin{figure}[tb]
    \centering
    \includegraphics[width=\linewidth]{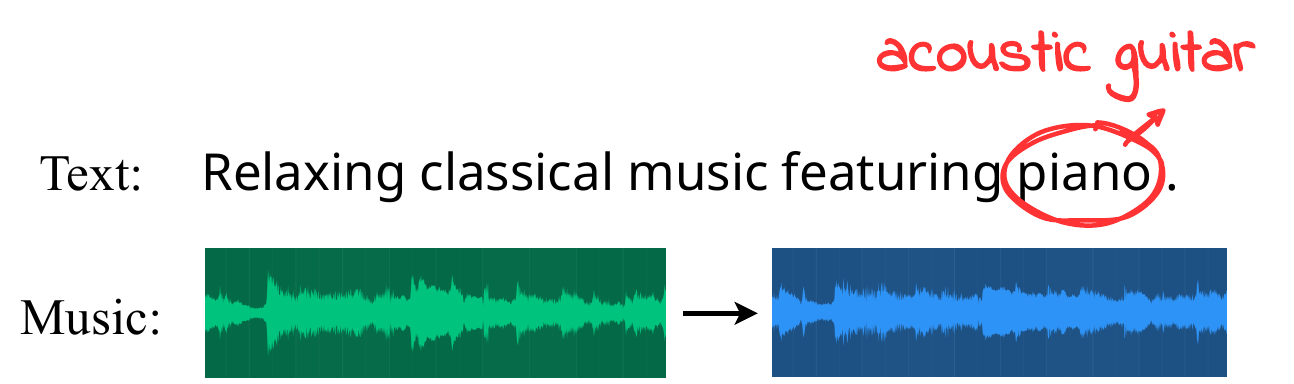}
    \caption{Text-to-music editing with MusicMagus. The edit from ``piano" to ``acoustic guitar" in the text prompt directly alters the corresponding musical attribute, while leaving others unchanged.}
    \label{fig:teaser}
\end{figure}

In this work, we introduce MusicMagus, which focuses on text-based \textit{intra-stem} music editing. Leveraging the inherent capabilities of pre-trained diffusion models, MusicMagus is able to perform zero-shot editing without requiring additional training pairs. As illustrated in Figure~\ref{fig:teaser}, we utilise word swapping to direct the editing process. This operation is implemented as a manipulation within the diffusion model's semantic space. Recognizing the sensitivity of the diffusion process, where minor alterations can propagate significant changes, we employ an additional constraint to ensure that the resultant music maintains the structural integrity and stylistic coherence of the original music. 

Although we mainly focus on the editing of music clips generated from diffusion models, we also discuss how to edit real-world music audio by the Denoising Diffusion Implicit Model (DDIM) inversion~\cite{ddim} technique.

In summary, our main contributions are as follows:

\begin{enumerate}[itemsep=0pt, parsep=0pt]
    \item We propose a flexible and user-friendly text-to-music editing method using word swapping.
    \item We contribute MusicMagus, a system capable of zero-shot music editing on diverse tasks without any dependence on \textit{paired} training data.
    \item Comparative experiments validate that MusicMagus outperforms existing zero-shot methods and some supervised approaches in critical tasks such as style and timbre transformation. \footnote{Work done during Yixiao’s internship at Sony AI.} \footnote{Code and demo: \url{https://bit.ly/musicmagus-demo}.}
\end{enumerate}

\section{Related Work}

\subsection{Text-to-Music Generation}

Text-to-music generation models in the audio domain broadly fall into two categories: autoregressive (AR) models, primarily language model (LM) based, operating on discrete audio representations, and diffusion-based models working with continuous latent representations~\cite{ddpm}. AR models like MusicLM~\cite{musiclm}, MeLoDy~\cite{melody} and MusicGen~\cite{musicgen} excel in creating longer and higher-quality audio sequences but are limited by higher inference times, which can be challenging for interactive applications such as music generation and editing. Conversely, diffusion models, including M{\"o}usai~\cite{mousai}, AudioLDM 2~\cite{audioldm2}, Jen-1~\cite{jen1}, and Tango~\cite{tango}, offer advantages in parallel decoding but require numerous diffusion steps for high-quality output, and often struggle with generating longer audio sequences. Recently, MagNet~\cite{magnet} offers a novel, hybrid approach to music generation. Combining the best of AR and diffusion models, it starts with autoregressive sequence generation and finishes with parallel decoding. This method effectively balances quality and efficiency in music production.

There is also a growing emphasis on controllability in text-to-music generation models. Coco-mulla~\cite{cocomulla} utilizes drum patterns and chord progressions, enhancing MusicGen's conditional music generation capabilities. Similarly, Music ControlNet~\cite{musicontrolnet} and DITTO~\cite{ditto} apply multiple controls over a pretrained diffusion model for tailored music creation. Mustango~\cite{mustango} integrates metadata control within the diffusion-based TANGO~\cite{tango} framework; whereas Jen-1 Composer~\cite{jen1composer} and StemGen~\cite{stemgen} generate new stems conditioned on existing stems, thus capitalizing on pre-existing musical elements for generation.

\subsection{Text-to-Music Editing}

Text-to-music editing encompasses two distinct types of operations: inter-stem and intra-stem editing. Inter-stem editing refers to operations conducted on one stem (such as adding or removing stems) that are conditioned on another stem, whereas intra-stem editing involves modifications within the stem itself, such as adjusting the instrument, genre, or mood.

Compared to text-based image editing ~\cite{prompt2prompt,pix2pix,instructimagen}, research on text-to-music editing is relatively limited.  Models like InstructME~\cite{instructme} and M$^2$UGen~\cite{m2ugen} demonstrate capabilities in both inter-stem and intra-stem editing, allowing for structural changes and detailed modifications within stems, but they often require extra training and specific data. Loop Copilot~\cite{loopcopilot}, an AI agent, employs a combination of existing models to facilitate compositional editing, yet it does so without altering the fundamental architecture or interface of the original models. Furthermore, the tag-conditioned timbre~\cite{transplayer} and style~\cite{position} transfer task can be seen as a precursor to this task, however labels are not as flexible as text.

In contrast, our model introduces a novel intra-stem editing approach. While it also operates without additional training, our approach distinctively utilizes the latent capacities of pre-trained diffusion-based models. This method enables efficient text-to-music editing, leveraging existing model structures without necessitating their combination or alteration.

Concurrently, a recent work~\cite{ddpm-edit} also attempts the similar task of text-based audio editing with diffusion models. We encourage the readers to refer to the work for more details.

\section{Background}

MusicMagus utilizes a pretrained diffusion model~\cite{ddpm} for text-to-music editing, eliminating the need for additional training. Specifically, we use a pretrained AudioLDM 2 model~\cite{audioldm2} as the backbone model, since it is the most accessible text-to-music diffusion model with publicly available weights. AudioLDM 2 employs a variational autoencoder (VAE)~\cite{vae} to compress a music audio spectrogram into a latent low-dimensional space. It then trains a latent diffusion model (LDM) on this latent space to generate new samples from Gaussian noise conditioned on text inputs. During generation, the LDM takes a condition \( y \), generates a latent variable \( z_0 \), and uses the VAE decoder to produce the music spectrogram \( x \), which can be then converted into a waveform using an external vocoder, such as HiFi-GAN~\cite{hifigan}.

During training, the LDM performs a forward diffusion process, which is defined as a Markov chain that gradually adds Gaussian noise to the latent representation of the data over \( T \) steps. This process can be represented as:
\begin{equation}
    z_{t} = \sqrt{\alpha_t} z_{t-1} + \sqrt{1 - \alpha_t} \epsilon, \quad \epsilon \sim \mathcal{N}(0, I),
\end{equation}
where \( t = 1, 2, \ldots, T \), \( z_t \) is the latent variable at step \( t \), \( \alpha_t \) is a variance schedule for the noise, and \( \epsilon \) is a noise vector drawn from a standard Gaussian distribution. The process starts with \( z_0 \) being the initial latent representation of the data and ends with \( z_t \) being a sample from the Gaussian noise distribution.

The inference process in LDMs is the reverse of the forward process. It starts with a sample from the Gaussian noise distribution \( z_t \) and aims to recover the original data representation \( z_0 \). This is achieved by a series of denoising steps that can be described by the following formulation:
\begin{equation}
    z_{t-1} = \frac{1}{\sqrt{\alpha_t}} \left( z_t - \frac{1 - \alpha_t}{\sqrt{1 - \bar{\alpha}_t}} \epsilon_\theta(z_t, t) \right) + \sigma_t\epsilon, \quad \epsilon \sim \mathcal{N}(0, I) 
    \label{eq:denoise}
\end{equation}
where \( \bar{\alpha}_t = \prod_{s=1}^{t} \alpha_s \) and \( \epsilon_\theta(z_t, t) \) is a neural network that predicts the noise added at step \( t \), and $sigma_t$ represents the standard deviation of the noise added. The network \( \epsilon_\theta \) is trained to minimize the difference between the predicted noise and the actual noise added during the forward process.

For simplicity, we denote the formula (\ref{eq:denoise}) as:

\begin{equation}
    z_{t-1} = \text{Denoise}(z_t, \epsilon_\theta, t).
\end{equation}

To decrease computational demands, denoising diffusion implicit models (DDIM)~\cite{ddim} introduced a modified approach which enables significantly fewer sampling steps (e.g., between 50 and 100, whereas DDPMs usually have 1000 steps) during inference, while having a negligible effect on the quality of the generated output.

\section{Method}

To illustrate our idea, we refer to the example in Figure~\ref{fig:teaser}. Initially, a music clip, denoted as $x$, is generated from the text prompt ``Relaxing classical music featuring piano", which we refer to as $y$. The next step involves altering this text prompt by substituting ``piano" with ``acoustic guitar", thereby creating a new prompt $y'$. Our aim is to produce a revised music piece $x'$, where only the specified attribute is changed, while maintaining all other aspects.

The explanation of our idea is twofold. In Section~\ref{sec:method_1}, we detail the method for altering the text prompt in the semantic domain. Subsequently, in Section~\ref{sec:method_2}, we discuss our approach to enforce suitable constraints over the cross-attention map during diffusion to preserve the integrity of the remaining elements of the music.


\subsection{Finding Editing Direction}\label{sec:method_1}

In this section, we introduce a strategy to calculate a difference ($\Delta$) vector in the latent space to guide the editing direction. This method is chosen over direct word swapping as it better preserves semantic coherence and contextual relevance, especially in cases of varying phrase lengths and complex content alterations. We will further explain it in Section~\ref{sec:method_2}; besides, previous research finds that similar operations can facilitate a more robust edit, especially when the keywords subject to modification are sparsely represented in the training dataset~\cite{pix2pix}.


We first introduce the text embedding method in Audio\-LDM 2. AudioLDM 2 uses a two-branch text encoder to embed the text prompt $y$ to two embeddings: $E = \{E_\text{T5}, E_\text{GPT}\}$, where $E_\text{T5}$ encodes the sentence-level representation, and $E_\text{GPT}$ captures the more fine-grained semantic information inside $y$.

First, the FLAN-T5~\cite{flant5} encoder, utilizing a T5 model~\cite{t5}, encodes \( y \) into a feature vector \( E_\text{T5} \in \mathbb{R}^{L \times 1024} \), where $L$ represents the sentence length. In parallel, the CLAP~\cite{clap} text encoder leverages a RoBERTa~\cite{roberta} model to transform \( y \) into a flattened vector \( E_\text{CLAP} \in \mathbb{R}^{1 \times 512} \): 

\begin{equation}
\left\{
\begin{aligned}
E_\text{T5} &= \text{T5}(y), \\
E_\text{CLAP} &= \text{CLAP}(y).
\end{aligned}
\right.
\end{equation}

Then, $E_\text{T5}$ and $E_\text{CLAP}$ are  linearly projected to \( P \in \mathbb{R}^{768} \). A GPT-2 model, pre-trained on an AudioMAE~\cite{audioMAE}, is then employed to auto-regressively generate 8 new tokens \( E_\text{GPT} \in \mathbb{R}^{8 \times 768} \):

\begin{equation}
E_\text{GPT} = \text{GPT-2}(\text{Proj}(E_\text{T5}, E_\text{CLAP})).
\end{equation}

The LDM takes both $E_\text{T5}$ and $E_\text{GPT}$ as input in the diffusion process:

\begin{align}
\epsilon_\theta &= \epsilon_\theta(z_t, E, t), \\
z_{t-1} &= \text{Denoise}(z_t, \epsilon_\theta, E, t).
\end{align}

Similarly, the new prompt $y'$ can be encoded to $E' = \{E'_\text{T5}, E'_\text{GPT}\}$. Our goal is to find $E^\text{edit} = \{E^\text{edit}_\text{T5}, E^\text{edit}_\text{GPT}\}$.

\begin{figure}[tb]
    \centering
    \includegraphics[width=\linewidth]{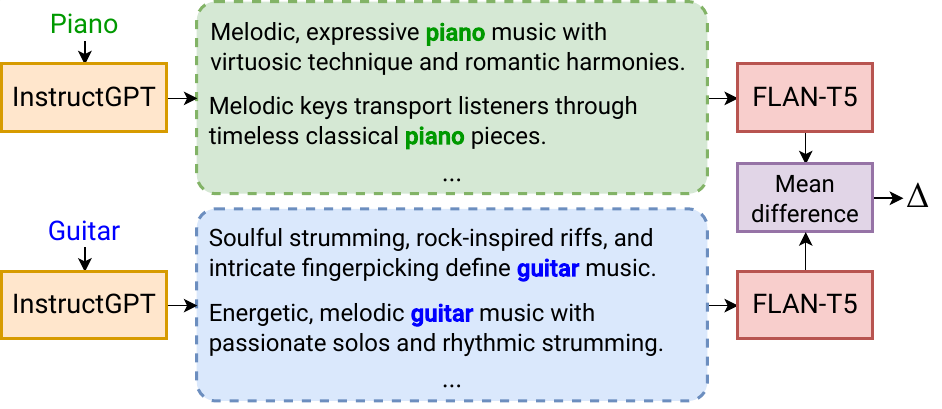}
    \caption{The pipeline of finding the editing direction $\Delta$. We first use InstructGPT to generate a large number of captions and then calculate the mean difference between the two embedding sets.}
    \label{fig:delta}
\end{figure}

We use the following method to find the editing vector $\Delta$, as shown in Figure~\ref{fig:delta}:

\begin{enumerate}[itemsep=0pt, parsep=0pt]
    \item We first generate a multitude of music-related captions using a pretrained InstructGPT model~\cite{instructGPT}. These captions are designed to contain the original and new keywords.
    \item Subsequently, we input these two sets of captions into the FLAN-T5 encoder and compute the mean embeddings for each set of encoded vectors.
    \item The final step is calculating the difference between these two mean embeddings, which is then employed as the vector for the editing direction $\Delta$.
\end{enumerate}

We employ different strategies to edit $E_\text{T5}$ and $E_\text{GPT}$. For $E_\text{T5}$, the edited embedding is:

\begin{equation}
    E^\text{edit}_\text{T5} = E_\text{T5} + \Delta.
\end{equation}

The aforementioned editing method encounters challenges when applying $\Delta$ to \( E_{\text{GPT}} \). The core issue is that \( E_{\text{GPT}} \) is obtained through the GPT-2 model, where the addition of a \( \Delta \) to the embedding may not constitute a semantically valid operation. Consequently, in practical applications, we resort to using \( E^\text{edit}_{\text{GPT}} = E'_{\text{GPT}}\), which is derived directly from encoding the new prompt.

Finally, we have the edited embeddings:

\begin{equation}
    E^\text{edit} = \{E_\text{T5} + \Delta, E'_{\text{GPT}}\}.
\end{equation}

\subsection{Adding Constraints Over Cross-Attention}\label{sec:method_2}

\begin{figure*}[htbp]
    \centering
    \includegraphics[width=\linewidth]{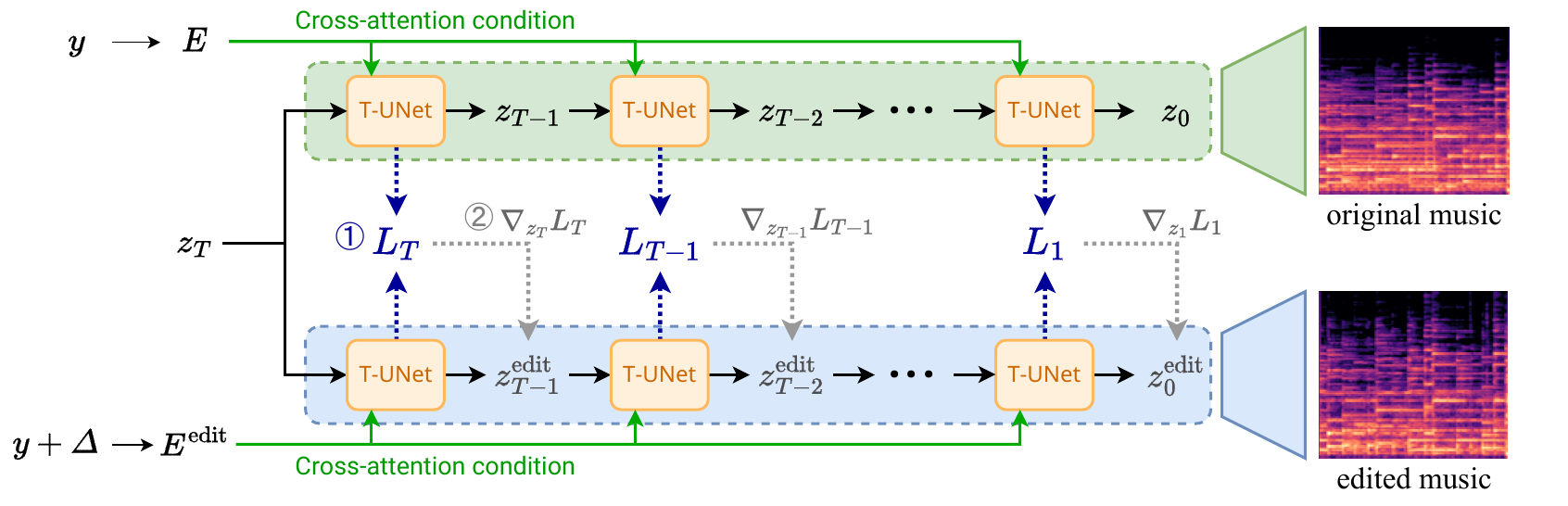}
    \caption{The workflow of the MusicMagus model. To constrain the diffusion model at timestep $t$, we need to: (1) calculate the L2 loss $L_t$ between the cross-attention map $M^\text{edit}_t$ and $M^\text{origin}_t$; (2) compute the gradient of $L_t$ with respect to $z_t$, and then perform a single-step optimization to update $\epsilon_\theta^\text{edit}$ of the diffusion model.}
    \label{fig:diagram}
\end{figure*}

Diffusion models exhibit inherent randomness in their generation output. By setting a fixed random seed and using the same text prompts, we can reproduce the same musical output. However, even minor variations in the text prompt can result in significantly different music clips. Previous studies have demonstrated that imposing external constraints on the cross-attention map between the text condition and the diffusion latent space enhances the consistency of the music generation, particularly for the remaining attributes that need to remain unchanged~\cite{prompt2prompt,pix2pix,plugandplay}. Building on this concept, we introduce a method designed to constrain the text-to-music diffusion model specifically for editing purposes.

To begin, we examine the acquisition of the cross-attention map. During the denoising process at timestep $t$, the model computes the cross-attention score between the encoded text $\{E_\text{T5}, E_\text{GPT}\}$ and the intermediate features of LDM $\epsilon_\theta$:

\begin{equation}
\begin{split}
\text{Attention}(Q, K, V) &= M \cdot V, \\
\text{where } M &= \text{Softmax}\left(\frac{QK^T}{\sqrt{d}}\right).
\end{split}
\end{equation}

In this context, $Q = W_Q\phi(z_t),~ K = W_kE,~ V = W_vE$ are defined, where $W = \{W_Q, W_K, W_V\}$ represents projection layers, and $E = \{E_\text{T5}, E_\text{GPT}\}$ are the text embeddings. AudioLDM 2 proposes the T-UNet architecture, which is distinct from the UNet architecture, to extract intermediate spatial features $\phi(x_t)$. T-UNet incorporates a transformer block after each encoder and decoder block's convolution operation, and the cross-attention occurs in the transformer block's final layer. The term $d$ denotes the dimension of the projected keys and queries.

As illustrated in Figure~\ref{fig:diagram}, to apply the editing, we first reconstruct the music $x$ with the original text embeddings $E$. We record the cross-attention maps for each timestep $t \in [1, T]$:

\begin{equation}
    M^\text{origin} = \{M^\text{origin}_1, ..., M^\text{origin}_T\}.
\end{equation}

\noindent Then we use the edited text embeddings $E^\text{edit}$ to generate an edited music clip. Similarly, at timestep $t$, we have a cross-attention map $M^\text{edit}_t$.

At each timestep $t$, we apply the constraint by calculating the $L_2$ loss between $M^\text{origin}_t$ and $M^\text{edit}_t$:

\begin{equation}
    L_t = \| M^\text{edit}_t - M^\text{origin}_t\|_2.
\end{equation}

\noindent We then compute the gradient $\nabla_{z_t}L_t$ and perform a single-step optimization with the step length $\alpha$:

\begin{equation}
    \epsilon^\text{edit}_\theta = \epsilon_\theta(z_t - \alpha\nabla_{z_t}L_t, E^\text{edit}, t).
\end{equation}

Subsequently, we execute the $t$-step denoising process using the updated $\epsilon^\text{edit}$:

\begin{equation}
    z_{t-1} = \text{Denoise}(z_t, \epsilon^\text{edit}_\theta, E^\text{edit}, t).
\end{equation}

\noindent This optimization is applied at every step until the denoising process is completed. Experimental results of the ablation studies validate that this constraint significantly enhances structural consistency during denoising.

To effectively utilize the cross-attention constraint, employing $\Delta$ for editing is essential. This method is crucial, especially when dealing with cases that involve substituting text of varying lengths, exemplified by replacing a shorter expression with a longer one (such as ``piano" $\rightarrow$ ``acoustic guitar"). Utilizing $\Delta$ maintains the uniformity of embedding lengths during the editing process. In contrast, techniques like word swapping can alter these lengths, leading to discrepancies between $M^\text{edit}$ and $M^\text{origin}$, and consequently, errors in calculating $L_t$. Furthermore, $\Delta$ facilitates the insertion of words at different sentence positions without disrupting the position-related cross-attention maps, ensuring the attention mechanism remains focused on the correct semantic context.

\section{Experiments}

In the domain of text-to-music editing, comprehensive model evaluation is inherently challenging due to the countless number of possible editing schemes. To address this, we focus on two key aspects: timbre transfer and style transfer, and compare our model's performance against established baselines in these areas. This comparison is conducted through both objective and subjective testing methodologies. 

\subsection{Baselines}

We benchmark our model against three distinct models in the field: AudioLDM 2~\cite{audioldm2}, Transplayer~\cite{transplayer}, and MusicGen~\cite{musicgen}. While our approach utilizes AudioLDM 2 as its backbone, AudioLDM 2 independently offers methods for both timbre and style transfer tasks, making it a relevant baseline.

\noindent\textbf{AudioLDM 2:} AudioLDM 2 is a diffusion-based model supporting unified speech, audio, and music generation at 16kHz. It follows the idea of AudioLDM and individually proposes a method for general audio style transfer. This is achieved through the interpolation of audio latents and subsequent denoising with a new prompt.

\noindent\textbf{Transplayer:} This state-of-the-art, diffusion-based model trained on POP909~\cite{pop909} and MAESTRO~\cite{maestro} dataset, specialising in timbre transfer at 16kHz. Unlike typical timbre transfer models that require training for each instrument pair, Transplayer is trained on multiple pairs, enabling versatile many-to-many timbre transfers.

\noindent\textbf{MusicGen:} A leading text-to-music generation model, MusicGen is a supervised model trained on a dataset of over 20,000 high-quality music pieces, generating 32kHz music. It uniquely allows for the inclusion of an extra melody condition, facilitating the style transfer task within the text-to-music generation process.

\subsection{Metrics}

We employ different metrics for subjective and objective experiments. For the subjective evaluation, we incorporate the following metrics, where OVL and REL are following~\cite{audiogen}:

\noindent\textbf{Overall Quality (OVL)} This metric is used to assess the overall music quality, encompassing aspects like sound clarity and musicality. It primarily evaluates whether the editing process enhances or diminishes the quality of the original music audio. The scoring for this metric ranges from 0 to 100.

\noindent\textbf{Relevance (REL)} REL measures the perceived semantic closeness between the edited music and the new text prompt. It is a subjective score, also ranging from 0 to 100.

\noindent\textbf{Structural Consistency (CON)} We define a new metric CON to evaluate the consistency of the pitch contour and structural aspects in the subjective test. Similar to the others, its scoring range is from 0 to 100.

The objective experiments utilize the following metrics:

\noindent\textbf{CLAP Similarity (CLAP)}~\cite{clap}: This metric assesses the semantic relevance between the edited music and the new text prompt. It utilizes a pretrained CLAP model, where a higher score indicates greater semantic similarity between the music and text, with scores ranging from 0 to 1. We implement it with the MuLaB library~\cite{mulab}.

\noindent\textbf{Chromagram Similarity (Chroma)}: We use this new metric to gauge the preservation of pitch contours and rhythm patterns in the music. It involves computing the cosine similarity between the chromagrams of the original and edited music. A higher score suggests better retention of the structure and pitch contour, with values also ranging from 0 to 1. We implement this metric with the librosa library~\cite{librosa}.

\subsection{Data Preparation}

\subsubsection{Objective Experiments}

For the timbre transfer task, we conducted a random selection of 60 music audio samples generated by AudioLDM 2, covering three specific word swapping pairs: (piano $\rightarrow$ organ), (viola $\rightarrow$ piano), and (piano $\rightarrow$ acoustic guitar). The primary rationale behind choosing these pairs is the limited range of instrument pairs supported by the Transplayer model. Given that the quality of music generated by AudioLDM 2 can vary, we implemented a quality-based filtering process. This entailed excluding any music samples that fell below a predefined quality threshold, continuing this selection process until the requisite number of suitable samples was attained.

Building upon the methodology established for timbre transfer, we applied a similar approach to the music style transfer task. Our selection encompassed a diverse range of style conversions, including (jazz $\rightarrow$ classical), (country $\rightarrow$ metal), (jazz $\rightarrow$ metal), and (jazz $\rightarrow$ rock). For each of these style pairs, we employed a random selection process, ultimately curating a dataset comprising 50 samples in total. 

We use a template to synthesize the text prompt: ``A \{\textit{mood}\} \{\textit{genre}\} music with \{\textit{timbre}\} performance.", where mood is randomly chosen from a fixed set of \{``upbeat", ``relaxing", ``peaceful"\}.

\subsubsection{Subjective Experiments}

For the subjective test, we randomly selected a subset of data points from the objective test dataset. Specifically, 8 data points were chosen for the timbre transfer task and 5 data points for the style transfer task. Each data point included results from both the baseline models and our ablation studies. The results are shown in Tables~\ref{tab:timbre_sub} and~\ref{tab:style_sub}. 

\subsection{Experimental Setup}

We choose the \textit{AudioLDM2-base} model~\footnote{\url{https://huggingface.co/cvssp/audioldm2}} as our backbone model. During inference, we configure the DDIM steps to 100, and generate 5-second audio clips at a sampling rate of 16kHz. A uniform gradient step length ($\alpha = 0.04$) is applied for both timbre transfer and style transfer tasks. All inference is performed on a single NVIDIA A100 GPU.

For the Transplayer model, we utilize the official pretrained checkpoint~\footnote{\url{https://github.com/Irislucent/TransPlayer}} without any modifications to its weights or code. As for MusicGen, we opt for the \textit{MusicGen-melody} checkpoint~\footnote{\url{https://huggingface.co/facebook/musicgen-melody}}, which has 1.5B parameters. To maintain consistency, all generated samples from these models are subsequently downsampled to 16kHz resolution.

\subsection{Results}

\subsubsection{Subjective Experiments}

We conducted a subjective listening test for both the timbre transfer and style transfer tasks. This test involved disseminating an online survey within the Music Information Retrieval (MIR) community and our broader research network, which resulted in the collection of 26 complete responses. The gender distribution of the participants was 19 males (76\%) and 6 females (24\%). Regarding musical experience, 5 participants (19.23\%) had less than 1 year of experience, 5 (19.23\%) had between 1 and 5 years, and the majority, 16 participants (61.54\%), had more than 5 years of experience. This subjective test was approved by the ethics committee of both Sony AI and Queen Mary University of London (QMERC20.565.DSEECS23.129).

The data presented in Table~\ref{tab:timbre_sub} reveals that our proposed model exhibits superior performance in the timbre transfer task when compared to two baseline models. Specifically, AudioLDM 2 demonstrates a notable limitation in transferring to novel semantics, resulting in edited samples that closely resemble the original ones. This is evident from its low Relevance (REL) score and high Consistency (CON) score. Contrary to expectations, the performance of Transplayer is consistently inferior, suggesting that its generalization capability may be inadequate for complex tasks such as many-to-many instrument timbre transfer in practical applications. Our model is the best on the average of altering semantic content and maintaining structural integrity. 

\begin{table}[tb]
\small
    \centering
    \begin{tabular}{l|l|ccc|c}
    \toprule
    \textbf{Model name} & \textbf{Type}    &  \textbf{REL}  & \textbf{OVL} & \textbf{CON} & \textbf{Avg.} \\
    \midrule
    AudioLDM 2 &   Zero-shot & 15.7 & 49.9 & \textbf{80.6} & 48.7\\
    Transplayer & Supervised & 28.3 & 28.9 & 34.6 & 30.6\\
    \midrule
    Ours \textit{w/o L2 \& $\Delta$} & Zero-shot & 78.0 & 61.6 & 50.4 & 63.3\\
    Ours \textit{w/o L2} & Zero-shot &\textbf{78.8}&\textbf{62.4}&51.3 & 64.2\\
    \midrule
    \textbf{Ours (final)} & Zero-shot &76.2&62.1&66.6 & \textbf{68.3}\\
    \bottomrule
    \end{tabular}
    \caption{The subjective evaluation results on the timbre transfer task.}
    \label{tab:timbre_sub}
\end{table}

Insights gleaned from our ablation study further elucidate these findings. The inclusion of the additional constraint significantly enhances performance in terms of Structure Consistency (CON), highlighting its role in bolstering structural coherence. However, the subjective experiments indicate no marked difference in Relevance (REL) scores between the methods. This observation aligns with expectations, since the primary objective of $\Delta$ usage is to ensure the consistency of the cross-attention maps, particularly during complex editing operations or in scenarios involving underrepresented words demonstrated in Section~\ref{sec:method_1}, which may not be fully reflected by the current subjective test settings.

We also evaluated our model's performance in the style transfer task, as detailed in Table~\ref{tab:style_sub}. Similar to the previous findings, our model demonstrates superior performance over the baseline models in this task as well.

\begin{table}[tb]
\small
    \centering
    \begin{tabular}{l|l|ccc|c}
    \toprule
    \textbf{Model name} & \textbf{Type}    &  \textbf{REL}  & \textbf{OVL} & \textbf{CON} & \textbf{Avg.} \\ 
    \midrule
    AudioLDM 2 &   Zero-shot & 19.8 & 53.2 & \textbf{84.2} & 52.4\\
    MusicGen & Supervised & 63.3 & \textbf{66.0} & 48.2 & 59.1\\
    \midrule
    Ours \textit{w/o L2 \& $\Delta$} & Zero-shot &69.2&56.9&58.9 & 61.7\\
    Ours \textit{w/o L2} & Zero-shot &\textbf{71.3}&53.8&55.0 & 60.0\\
    \midrule
    \textbf{Ours (final)} & Zero-shot & 65.7&57.8&65.6 & \textbf{63.1}\\
    \bottomrule
    \end{tabular}
    \caption{The subjective evaluation results on the style transfer task.}
    \label{tab:style_sub}
\end{table}

AudioLDM 2 exhibits notable limitations in style transfer, with its performance being generally unstable; MusicGen, despite its downsampled audio quality from 32KHz to 16kHz, retains a high level of audio quality, as indicated by its high Overall Quality (OVL) score. However, MusicGen struggles with precisely preserving the original melody in the style transfer process, particularly in maintaining polyphonic melodies, which introduces some instability in its outputs.

In contrast, our method not only changes the semantics but also keeps that the overall quality is not diminished, resulting in the best average score; it also maintains the structural integrity and pitch consistency, which are critical in music style transfer.

\subsubsection{Objective Experiments}

We compare the performance of our model and the zero-shot and supervised baselines. The results for the timbre transfer and style transfer tasks are shown in Tables~\ref{tab:timbre_obj} and~\ref{tab:style_obj}.

In the timbre transfer task (Table~\ref{tab:timbre_obj}), our model demonstrated enhanced performance in semantic transfer. The incorporation of a constraint on the cross-attention mechanism largely improved pitch and rhythm accuracy, reinforcing the insights obtained from the subjective experiments. These results substantiate the efficacy of our model in maintaining semantic integrity while facilitating timbre transfer results.

\begin{table}[tb]
\small
    \centering
    \begin{tabular}{l|l|cc|c}
    \toprule
    \textbf{Model name} & \textbf{Type}    &  \textbf{CLAP}  & \textbf{Chroma} & \textbf{Avg.}  \\
    \midrule
    AudioLDM 2 &   Zero-shot & 0.16 & 0.72 & 0.44 \\
    Transplayer & Supervised & 0.18 & 0.56 & 0.37 \\
    \midrule
    Ours \textit{w/o L2 \& $\Delta$} & Zero-shot  & 0.33 & 0.68 & 0.51\\
    Ours \textit{w/o L2} & Zero-shot & \textbf{0.34} & 0.69 & 0.52\\
    \midrule
    \textbf{Ours (final)} & Zero-shot & 0.33 & \textbf{0.76} & \textbf{0.55}\\
    \bottomrule
    \end{tabular}
    \caption{The objective evaluation results on the timbre transfer task.}
    \label{tab:timbre_obj}
\end{table}

Table~\ref{tab:style_obj} presents the findings for the style transfer task. Here, our model outperformed the baselines in terms of structural and pitch consistency. However, in terms of semantic transfer, the differences between our model and the baselines were less pronounced. This suggests that while our model excels in maintaining the structural and pitch elements during style transfer, the semantic changes are comparable to those achieved by the baseline models.

\begin{table}[tb]
\small
    \centering
    \begin{tabular}{l|l|cc|c}
    \toprule
    \textbf{Model name} & \textbf{Type}    &  \textbf{CLAP} & \textbf{Chroma} & \textbf{Avg.} \\
    \midrule
    AudioLDM 2 &   Zero-shot & 0.18 & \textbf{0.80} & \textbf{0.49}\\
    MusicGen & Supervised & \textbf{0.24} & 0.66 & 0.45\\
    \midrule
    Ours \textit{w/o L2 \& $\Delta$} & Zero-shot & 0.22 & 0.65 & 0.44\\
    Ours \textit{w/o L2} & Zero-shot & 0.22 & 0.67 & 0.45\\
    \midrule
    \textbf{Ours (final)} & Zero-shot & 0.21 & 0.77 & \textbf{0.49}\\
    \bottomrule
    \end{tabular}
    \caption{The objective evaluation results on the style transfer task.}
    \label{tab:style_obj}
\end{table}

\section{Discussion}

\subsection{Real Music Audio Editing}\label{sec:real}

\begin{figure*}[htbp]
    \centering
    \includegraphics[width=0.82\linewidth]{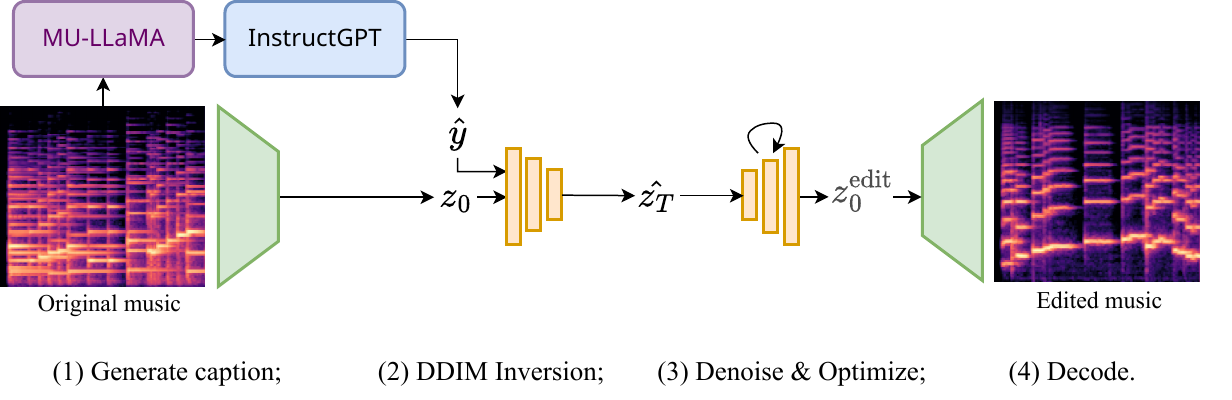}
    \caption{The diagram of the real music audio editing pipeline using MusicMagus with DDIM inversion and diffusion model editing.}
    \label{fig:inversion}
\end{figure*}

MusicMagus offers capabilities for editing real-world music audio, although it is noted that the performance may not match the editing of synthesized music audio generated from diffusion models. We begin with the DDIM inversion to estimate the latent representation $\hat{z_T}$ of a given real music audio $x$. This step is crucial to facilitate editing with the diffusion model, as depicted in Figure~\ref{fig:inversion}.

The inversion requires a corresponding text prompt $\hat{y}$, which is initially generated by a pretrained music captioning model, MU-LLaMA~\cite{mullama}. Due to the discrepancy between the text prompt distributions of AudioLDM 2 and MU-LLaMA, the InstructGPT model is employed to refine the generated captions, aligning them more closely with AudioLDM 2's distribution. This refinement includes condensing the caption into a single, concise sentence and emphasizing essential characteristics such as the key instruments, mood, and genre.

DDIM inversion, while effective, is not a perfect reconstruction method. It faces a trade-off between the editability of the estimated latent $\hat{z_T}$ and its reconstruction fidelity~\cite{prompt2prompt}. A balance is sought by selecting an intermediate value for classifier-free guidance, set to 1. Additionally, the diffusion latent is typically modeled as Gaussian noise. To mitigate auto-correlation that may arise during inversion, we adopt a strategy from Parmar et al.~\cite{pix2pix}, introducing autocorrelation regularization to diminish its impact, thereby enhancing the estimation of $\hat{z_T}$.

Subsequent to obtaining the estimated latent $\hat{z_T}$, the caption $\hat{y}$ is edited, and the MusicMagus editing algorithm is applied within the diffusion model framework to produce the edited music audio~\footnote{We provide listening samples at the demo page.}.

\subsection{Limitations}

The current implementation of MusicMagus, while effective, is built upon the AudioLDM 2 model, which is not without its constraints. One significant limitation is the model's challenge in generating multi-instrument music when such complexity is specified. This inherently restricts the scope of creative expression and diversity that the model can offer. The performance of AudioLDM 2 was not enhanced in our approach, which is an aspect we aim to address moving forward.

Moreover, our zero-shot method exhibits instability, as evidenced by a notable number of failure cases. These failures are often due to unsuccessful application of the delta and word-swapping techniques, highlighting an area ripe for improvement. Currently, the scope of alterations we can apply to the music is somewhat modest; our system struggles to introduce substantial changes, such as adding or removing an instrument, adding sound effects, etc., without compromising the overall structure and quality of the audio.


Another factor that confines our system is the inherent limitations of the base model itself. For instance, the diffusion process struggles with generating very long sequences, which in turn limits the practical applications of our model. Addressing this limitation could potentially open up new domains where longer sequence generation is essential.

Lastly, the audio quality, currently capped by the 16kHz sampling rate, is another significant limitation, often resulting in artifacts that can detract from the listener's experience. Enhancing the audio fidelity is an important step that will bring us closer to a model that can produce professional-grade audio, which is crucial for both consumer applications and artistic endeavors. The pursuit of higher audio quality and the reduction of artifacts are critical goals for our future work.

\section{Conclusion}

In conclusion, our research contributes a novel text-to-music editing framework that effectively manipulates selected musical aspects, such as timbre and style, without altering the remaining parts. Our method distinguishes itself by its compatibility with current diffusion models and its operational simplicity, not necessitating further training protocols. The empirical evidence from our studies confirms that our method advances the state-of-the-art, delivering enhanced performance in style and timbre transfer.

Although we have identified areas for improvement, such as the model's ability to handle complex multi-instrument compositions and the stability of zero-shot methods, these challenges provide a clear trajectory for our ongoing research. By incrementally refining the underlying model and expanding the editing capabilities, we aim to push the boundaries of automated music generation and editing further. The ultimate goal is to refine the underlying model, enabling the generation and editing of high-fidelity, nuanced, and diverse musical compositions with simple and intuitive human input while maximizing creative expressiveness.

\section*{Ethical Statement}

The subjective tests are approved by the ethics committee of both Sony AI and Queen Mary University of London (QMERC20.565.DSEECS23.129).

\section*{Acknowledgments}

We want to thank Dr. Woosung Choi, Zineb Lahrichi, Liwei Lin and Liming Kuang for their contribution to this work. Yixiao Zhang is a research student at the UKRI Centre for Doctoral Training in Artificial Intelligence and Music, supported jointly by the China Scholarship Council, Queen Mary University of London and Apple Inc.


\bibliographystyle{named}
{\small \bibliography{ijcai23}}

\end{document}